\documentclass[journal,12pt,onecolumn,draftclsnofoot,]{IEEEtran}
\usepackage{amsmath,amsfonts,amssymb}
\usepackage{algorithmic}
\usepackage{algorithm}
\usepackage{array}
\usepackage[caption=false,font=normalsize,labelfont=sf,textfont=sf]{subfig}
\usepackage{textcomp}
\usepackage{stfloats}
\usepackage{url}
\usepackage{verbatim}
\usepackage{graphicx}
\usepackage{ntheorem}
\usepackage[algo2e]{algorithm2e} 
\usepackage{amsmath}
\usepackage{cite}
\usepackage{amsfonts}
\usepackage{lipsum}
\usepackage{mathtools}
\usepackage{cuted}
\usepackage{setspace}
\usepackage{multicol}

\usepackage{booktabs, multirow}
\usepackage{siunitx}

\usepackage{kantlipsum}
\usepackage{footnote}
\usepackage{bm}
\usepackage{float}

\makeatletter

\newcommand{\Rmnum}[1]{\expandafter\@slowromancap\romannumeral #1@}

\begin{document}

\title{Joint AP-UE Association and Power Factor Optimization for Distributed Massive MIMO}

\author{\IEEEauthorblockN{Mohd Saif Ali Khan*, Samar Agnihotri* and Karthik R.M.$^+$}\\
\IEEEauthorblockA{*School of Computing \& EE, Indian Institute of Technology Mandi, HP, India}\\
\IEEEauthorblockA{$^+$Ericsson India Global Services Pvt. Ltd., Chennai, TN, India}\\
Email: saifalikhan00100@gmail.com, samar.agnihotri@gmail.com, r.m.karthik@gmail.com
%
}




\maketitle

\begin{abstract}
The uplink sum-throughput of distributed massive multiple-input-multiple-output (mMIMO) networks depends majorly on Access point (AP)-User Equipment (UE) association and power control. The AP-UE association and power control both are important problems in their own right in distributed mMIMO networks to improve scalability and reduce front-haul load of the network, and to enhance the system performance by mitigating the interference and boosting the desired signals, respectively. Unlike previous studies, which focused primarily on addressing these two problems separately, this work addresses the uplink sum-throughput maximization problem in distributed mMIMO networks by solving the joint AP-UE association and power control problem, while maintaining Quality-of-Service (QoS) requirements for each UE. To improve scalability, we present an \textit{l}1-penalty function that delicately balances the trade-off between spectral efficiency (SE) and front-haul signaling load. Our proposed methodology leverages fractional programming, Lagrangian dual formation, and penalty functions to provide an elegant and effective iterative solution with guaranteed convergence. Extensive numerical simulations validate the efficacy of the proposed technique for maximizing sum-throughput while considering the joint AP-UE association and power control problem, demonstrating its superiority over approaches that address these problems individually. Furthermore, the results show that the introduced penalty function can help us effectively control the maximum front-haul load. 
\end{abstract}


\section{Introduction}
\IEEEPARstart{T}{he} distributed mMIMO offers a new paradigm that builds on the advantages of mMIMO in cellular networks, yet addresses the accompanying challenges \cite{ngo2017cell,ngo2017total}. Unlike traditional cellular networks, distributed mMIMO deploys a large number of access points (APs) throughout the coverage region without any cell boundaries. This departure from typical cell definition eliminates inter-cell interference, a persistent problem in cellular networks. Furthermore, UEs' closeness to these distributed APs may substantially enhance the coverage probability, improving overall QoS for users. 

In the original notion of distributed mMIMO \cite{ngo2017cell}, in the absence of traditional cell boundaries, it is assumed that all APs simultaneously serve all UEs over the same time and frequency resources. This assumption makes the system impractical, especially when the number of UEs increases dramatically, leading to substantial increase in the computational overhead and front-haul signaling load associated with Channel State Information (CSI) and data transfer \cite{bjornson2020scalable}. This leads to the introduction of scalable distributed mMIMO, as in \cite{bjornson2020scalable}, where each UE is served by a subset of APs. This is justified as  more than 95\% received  signal strength is concentrated in a few nearby APs.  Thereby, the AP-UE association is one of the important aspects of distributed mMIMO networks to reduce the computation costs and front-haul load. Power control is another  important aspect to achieve efficient utilization of resources in any communication network. Beyond its role in enhancing system energy efficiency, as emphasized in  \cite{ngo2017total}, efficient power control plays a crucial role in pro-actively minimizing interference, thereby improving the spectral efficiency (SE). Thus, the implementation of appropriate power control measures is highly desirable. So far these AP-UE association and power control problems have been handled separately. However, we argue and establish that jointly addressing these two problems is more effective in mitigating interference, boosting signal strength, and enhancing achievable throughput.   

A significant amount of work has been undertaken to address power control and AP-UE association problems in distributed mMIMO systems. However, it is worth noting that much of it tends to focus entirely on either power control \cite{ngo2017cell,ngo2017total,nikbakht2020uplink,bjornson2020scalable}, or AP-UE association \cite{bjornson2020scalable,buzzi2019user,ghiasi2022energy}. In \cite{ngo2017total} and \cite{vu2020joint}, the authors focus on downlink systems that offer inherent coupling between AP-UE association and power control making the joint analysis less complex in contrast to the uplink systems, which lack such natural coupling between the two. This highlights the importance of customized methodologies for extending joint optimization techniques from downlink to uplink distributed  mMIMO systems.
While \cite{guenach2021deep} introduces a deep learning-based joint approach for power control and AP-UE association in uplink scenarios, it has high computational cost of extensive training and lacks the explainability associated with traditional models.
In \cite{ngo2018performance} and \cite{guenach2020joint}, a novel approach is adopted by formulating an optimization problem from the perspective of maximizing the minimum Signal-to-Interference-Noise ratio for joint power control and AP-UE association weights for the uplink distributed mMIMO. However, there is an inherent difficulty in allocating APs to UEs within the formulated optimization problem. The solutions obtained provide the weights of the AP-UE association matrix, prompting the authors to propose an additional algorithm for AP-UE scheduling. 

In this work, we address the challenging nature of uplink distributed mMIMO systems by formulating a mixed-integer non-convex sum-throughput maximization problem that incorporates joint AP-UE association and power control, subject to the QoS requirements of each UE in terms of the SE. To navigate the trade-off between the SE and the front-haul load, we introduce an \textit{l}1-penalty approach \cite{tsuruoka2009stochastic}. This strategy allows us to control the trade-off between the SE and the front-haul load in the network as per the operational requirements. Our formulation incorporates an important consideration for preserving QoS, a minimum threshold of the SE for each UE, thus offering equitable connectivity experience for all users.  The optimization problem is solved by using an iterative approach with guaranteed convergence to a solution, by leveraging the fractional programming and the Lagrangian dual formulations.

\textit{Organization:} The paper is organised as follows: Section \ref{system_model} presents the system model outlining the key components and assumptions used in this study and subsection \ref{problem_formulation} details the problem formulation. Section \ref{solution_analysis} introduces our proposed solution and establishes its convergence. In Section \ref{simulations}, we numerically evaluate the performance of the proposed scheme and establish its effectiveness with respect to existing schemes. Finally, Section \ref{conclusion} concludes the paper by summarizing our work and discussing its future directions.

\section{System Model}
\label{system_model}
We consider a system consisting of $T$ single antenna UEs and $M$  multi-antenna APs, where each AP has $A$ antennas such that $T<< MA$. The APs and UEs are distributed uniformly over the given coverage area and APs are jointly serving all users over the same time and frequency resources. In order to help in the signal processing and coordination among APs, they are connected with a CPU via a front-haul connection. We consider the TDD mode of operation where the AP-UE channel vector  is estimated using the uplink transmission. To model our channel, we have considered the block fading model with a coherence block of length $L_c$. Let $L_p$ be the length of orthogonal pilots such that $L_p$ length of coherence block is used for uplink pilot training. The Rayleigh fading coefficient representing the small-scale fading is denoted by $\textbf{h}_{mt} \in \mathbb{C}^{\textit{A}\times 1 }$, whereas the large scale-fading coefficient (LSFC) that incorporates both path-loss and shadowing effects is denoted by $\beta_{mt}$. The channel vector $\textbf{g}_{mt} \in \mathbb{C}^{\textit{A}\times 1 }$, that incorporates both the small-scale fading as well as large scale fading, is defined as $\textbf{g}_{mt} = \beta_{mt}^{1/2}\textbf{h}_{mt}$.
We assume that  $\textbf{h}_{mt}{\sim} \mathcal{N}_{\mathbb{C}}(0,1)$ is independent and identically distributed (i.i.d.) random vector. 

\subsection{Uplink Pilot Training}
We assume that during uplink channel estimation, each UE $t \in \mathcal{T} \overset{\Delta}{=} \{1,..., T\} $ transmits the pilot sequence $\sqrt{p_pL_p}\mathbf{\Psi}_t \in \mathbb{C}^{L_p\times 1 }$, such that  $||\mathbf{\Psi}_t||^2 = 1$, where $p_p$ represents the normalized signal-to-noise power. The received signal $\textbf{y}^{pilot}_{m} \in \mathbb{C}^{\textit{A}\times L_p }$ at the $m^{th}$ AP is given by :
\begin{align*}
 \textbf{y}^{pilot}_{m} = \sqrt{p_pL_p}\sum_{t\in \mathcal{T}}\textbf{g}_{mt}\mathbf{\Psi}_{t}^{\mathit{H}} + \textbf{n}_m, 
\end{align*}
where $\textbf{n}_m \in \mathbb{C}^{\textit{A}\times L_p }$ is the complex additive white Gaussian noise matrix of the AP $m$ consists of complex i.i.d. elements. The MMSE estimate $\hat{\textbf{g}}_{mt}\in \mathbb{C}^{\textit{A}\times 1 }$ of $\textbf{g}_{mt}$ is given by \cite{ngo2017cell}:
\begin{align*}
\hat{\textbf{g}}_{mt} = \frac{\sqrt{p_pL_p}\beta_{mk}}{\sum_{t'=1}^{T}p_pL_p\beta_{mt'}\left| \Psi^{H}_{t}\Psi_{t'} \right|^2 +1}\textbf{y}^{pilot}_{m}\mathbf{\Psi}_t.
\end{align*}
The mean-square of the estimate $\hat{\textbf{g}}_{mt}$ for any antenna element given by, as in \cite{ngo2017cell}:
\begin{align*}
\gamma_{mt} = \left(\frac{p_pL_p\beta_{mt}^2}{\sum_{t'=1}^{T}p_pL_p\beta_{mt'}\left| \Psi^{H}_{t}\Psi_{t'} \right|^2 +1}\right).
\end{align*}

\subsection{Uplink Data Transmission}
Let $x_t$ be the uplink payload signal transmitted from the UE $t$, such that $\mathbb{E}\{|x_t|^2\}=1$. The uplink normalized signal-to-noise power, additive noise and power control coefficient for UE $t$ are denoted by $p_u$, $\textbf{n}_m {\sim} \mathcal{N}_{\mathbb{C}}(0,\textbf{I}_A)$ and $\eta^{u}_{t}$ respectively. The uplink signal received at the $m^{th}$ AP is given by:
\begin{align*}
 \textbf{y}^{u}_{m} = \sqrt{p_u}\sum_{t\in \mathcal{T}}\textbf{g}_{mt}\sqrt{\eta^{u}_{t}} x_t + \textbf{n}_m. 
\end{align*}
To detect the signal of the UE $t$ in a centralised manner, the AP $m$ selects the received combining vector $\textbf{v}_{mt} \in \mathbb{C}^{\textit{A}\times 1}$ locally and sends ${\textbf{v}}_{mt}^{H}\textbf{y}^{u}_{m}$ to the CPU. The received signal at the CPU for the UE $t$ is, as in \cite{ngo2017cell}:
\begin{align*}
 {{y}}^{u}_{t} =\sum_{m\in \mathcal{M}}{\textbf{v}}_{mt}^{{H}} \textbf{y}^{u}_{m}, 
\end{align*}
where $\mathcal{M} \overset{\Delta}{=} \{1,...,M\} $ is the set of all APs. Following the approach as in \cite{ngo2017cell} i.e., using the MR combining such that ${\textbf{v}}_{mt} = d_{mt}{\textbf{g}}_{mt}$, the closed form of the SE is given by:
\begin{align}
\label{eq_11}
& \text{SE}^{u}_t = w \log_2 \left(1+\Gamma^{*}_t\right), 
\end{align}
where $d_{mt}$ is the indicator variable indicating the association between the UE $t$ and the AP $m$, such that $d_{mt}=1$ if the AP $m$ serves the UE $t$, otherwise $d_{mt}=0$; and $\Gamma^{*}_t$ is defined as:
\begin{align}
\Gamma^{*}_t \overset{\Delta}  {=} \frac{A^2p_u\eta_{t}^{u}\left(\sum\limits_{m\in \mathcal{M}}d_{mt} \gamma_{mt}\right)^2}{{I}_t}, \label{eq_11_1}
\end{align}
with
\begin{align}
\begin{split}
 {I}_t \overset{\Delta}  {=} \sum\limits_{t'=1,t'\neq t}^{T}\!\!\!\!\!A^2p_u\eta_{t'}^{u}\left|\Psi^{H}_{t}\Psi_{t'} \right|^2\left(\sum\limits_{m\in \mathcal{M}}\!\!\!d_{mt}\gamma_{mt}\frac{\beta_{mt'}}{\beta_{mt}}\right)^2    +\sum\limits_{t'=1}^{T}\!\!\!Ap_u\eta_{t'}^{u}\!\!\!\sum\limits_{m\in \mathcal{M}}\!\!\!d_{mt}\gamma_{mt}\beta_{mt'} +\!\!\! \sum\limits_{m\in \mathcal{M}}\!\!\!Ad_{mt}\gamma_{mt}, \label{eq_12}
\end{split}
\end{align}
where the first, second and third terms in ${I}_t$ represent the interference caused by the pilot contamination, beamforming gain uncertainty, and noise, respectively.
 
\subsection{Problem Formulation}
\label{problem_formulation}
Our primary objective is to  maximize the total SE of all UEs. Furthermore, we ensure that each UE gets a minimum QoS. The minimum QoS for all the UEs is defined as the minimum SE that every UE gets. Further, to prevent the low SE providing APs from serving the UEs, we introduce the \textit{l}1-penalty in the objective function. This not only improves the energy efficiency, but also prevents the front-haul overloading\footnote{Computation of front-haul load in our framework is expressed mathematically as $\sum_{t \in \mathcal{T}} d_{mt}\text{SE}^{u}_t \ \forall m$. Additionally, the maximum front-haul is defined as the highest front-haul load among all APs. This metric provides insight into the peak load on the front-haul infrastructure across the entire network.}. This results in the following problem formulation:
\begin{subequations}
\label{eq : 13_m}
\begin{align} 
\label{eq : 13}
&\underset{\bm{\eta^{u}},\textbf{D}}{\max~} \sum\limits_{t=1}^{T}\left(\text{SE}^{u}_t  -\alpha ||{\textbf{D}_t}||_{1}\right) ,\\ 
& \!\!\!\!\!\!\!\!\!\!\!\!\!\!\!\!\!\!\!\!\!\!\!\!\!\!\!\!\!\!  \text{subject to :} \ d_{mt} \in \{0,1\}, \forall m \in \mathcal{M}, \  t \in \mathcal{T}, \label{eq : 13_1}\\
&0 \leq \eta^{u}_t \leq 1, \forall  t \in \mathcal{T}, \label{eq : 13_2}\\
&\text{SE}^{u}_t \geq \text{SE}^{\text{QoS}}_t, \forall  t \in \mathcal{T}, \label{eq : 13_3} \\
&\sum\limits_{m =1}^{M}d_{mt} \geq 1,  \forall  t \in \mathcal{T}, \label{eq : 13_4}
\end{align}
\end{subequations}
where $\bm{\eta^{u}} = [\eta^{u}_1, \eta^{u}_2,...\eta^{u}_T]^\intercal$ and $\textbf{D}$ is the AP-UE association matrix with each element as $d_{mt}$. $\textbf{D}_t$ represents the column vector of matrix of \textbf{D} with respect to the UE $t$. The regularization coefficient $\alpha \geq 0$ is the balancing factor between the front-haul load and the SE. $\text{SE}^{\text{QoS}}_t$ represents the minimum QoS requirement for the UE $t$. The constraint (\ref{eq : 13_1}) represents the binary  variable indicating the association between the UEs and APs. Constraint (\ref{eq : 13_2}) represents the uplink power control coefficient. Constraint (\ref{eq : 13_3}) is for the minimum QoS requirement and constraint (\ref{eq : 13_4}) specifies that at least one AP must be connected to the UE. The objective function in (\ref{eq : 13}) is a non-smooth and  mixed integer type non-convex function that makes the overall problem NP-hard. Finding the optimal solution of such mixed integer type problems requires high computation resources. In the following section we provide a heuristic solution of problem (\ref{eq : 13_m}) using a relaxation approach.

\section{Proposed Joint Solution using Relaxation Technique}
\label{solution_analysis}
Problem (\ref{eq : 13_m}) is a mixed-integer type problem which has a very high complexity, so we first convert the discrete binary constraint (\ref{eq : 13_1}) into a continuous one, resulting in:
\begin{align} 
\label{eq : 14_9}
 d_{mt} \in [0,1] , \forall m \in \mathcal{M}, \  t \in \mathcal{T}.
\end{align}
\textit{Remark 1 : The introduction of \textit{l}1-penalty in the objective function serves to induce the sparsity in the AP-UE association matrix $\textbf{D}$. By careful selection of  the trade-off factor $\alpha$, controlled-sparsity is achieved by limiting the number of APs that serve the UEs. Therefore, the matrix $\textbf{D}$ has non-zero values only for the chosen AP-UE pairs. After completion of  AP-UE association, all entries in $\textbf{D}$ undergo rounding to the nearest integer values i.e., zero and one, thus maintaining the feasibility of the optimization problem (\ref{eq : 13}). In the forthcoming results section, the SE with and without rounding-off of values in $\textbf{D}$ are compared to numerically show the effectiveness of our proposed approach.}

Then, the optimization problem (\ref{eq : 13}) can be formulated as:
\begin{subequations}
\label{eq : 14_m}
\begin{align} 
\label{eq : 14}
\underset{\bm{\eta^{u}},\textbf{D}}{\max~}& \ \ \sum\limits_{t=1}^{T}\left(w\text{log}_2 \left(\!1+\Gamma^{*}_t \right)  -\alpha ||{\textbf{D}_t}||_{1}\right) , \\
\text{subject to :} & \ \ d_{mt} \in [0,1], \forall m \in \mathcal{M}, \  t \in \mathcal{T}, \label{eq : 14_1}\\
&\ \ 0 \leq \eta^{u}_t \leq 1, \forall  t \in \mathcal{T}, \label{eq : 14_2}\\
&\ \ w\text{log}_2 \left(1+\Gamma^{*}_t \right) \geq \text{SE}^{\text{QoS}}_t, \forall  t \in \mathcal{T}, \label{eq : 14_3} \\
&\ \ \sum\limits_{m =1}^{M}d_{mt} \geq 1,  \forall  t \in \mathcal{T}. \label{eq : 14_4}
\end{align}
\end{subequations}
Considering the non-convexity due to the ratio term inside the log in our objective function (\ref{eq : 14}), we introduce the lower-bounding auxiliary variable $\Gamma_t$, resulting in the objective function with an additional constraint as follows:
\begin{subequations}
\label{eq : 16_m}
\begin{align} 
\underset{\bm{\eta^{u}},\textbf{D},\mathbf{\Gamma}}{\max~} & \ \ \sum\limits_{t=1}^{T}\left( w\text{log}_2 (1+\Gamma_t)  -\alpha ||{\textbf{D}_t}||_{1}\right), \label{eq : 16_1}\\
\text{subject to :}& \ \ \Gamma_t \leq  \Gamma^{*}_t , \  t \in \mathcal{T}, \label{eq : 16_2}
\end{align}
\end{subequations}
where $\mathbf{\Gamma} = [\Gamma_1,\Gamma_2,...\Gamma_T]^\intercal$. Taking the partial Lagrangian of objective function (\ref{eq : 16_1}) by considering constraint (\ref{eq : 16_2}) with respect to $\mathbf{\Gamma}$, we have:
\begin{align}
\label{eq : 17}
\begin{split}
L(\bm{\eta^{u}},\textbf{D},\mathbf{\Gamma},\bm{\lambda}) = & \sum\limits_{t=1}^{T}\left( w\text{log}_2 (1+\Gamma_t)  -\alpha ||{\textbf{D}_t}||_{1}\right) \\ & -\sum\limits_{t=1}^{T} \lambda_t\left(\Gamma_t - \Gamma^{*}_t    \right),   
\end{split}
\end{align}
where $\bm{\lambda} = [\lambda_1,\lambda_2,...\lambda_T]^\intercal$. 

After computing the stationary point of (\ref{eq : 17}) with respect to $\mathbf{\Gamma}$, we get the optimal value of $\lambda^{*}_t = \frac{w'}{1+{\Gamma^{*}_t}}$. where $w' = w/{\ln 2}$ with the optimal value of $\mathbf{\Gamma} =\mathbf{\Gamma^{*}} $, where $\mathbf{\Gamma^{*}} = [\Gamma^{*}_1,\Gamma^{*}_2,...\Gamma^{*}_T]^\intercal$. 
Substituting this $\lambda^{*}_t$ in  (\ref{eq : 17}) and using the Lagrangian dual transformation \cite{shen2018fractional}, the problem in (\ref{eq : 14_m}) can be reformulated as : 
\begin{subequations}
\label{eq : 18_m}
\begin{align}
\begin{split}
&\underset{\bm{\eta^{u}},\textbf{D}}{\max~}  \sum\limits_{t=1}^{T}\left( w\text{log}_2 (1+\Gamma_t)  -\alpha ||{\textbf{D}_t}||_{1}\right)   - \sum\limits_{t=1}^{T}\left(\Gamma_t - \frac{w'(1+{\Gamma}_t)A^2p_u\eta_{t}^{u}(\sum\limits_{m\in \mathcal{M}} d_{mt}\gamma_{mt})^2}{A^2p_u\eta_{t}^{u}(\sum\limits_{m\in \mathcal{M}}d_{mt} \gamma_{mt})^2+{I}_t}   \right), \label{eq : 18}\\
\end{split}
\end{align}
\begin{align}
\!\!\!\!\!\!\!\!\!\!\!\!\!\!\!\!\!\!\!\!\!\!\!\!\!\!\!\!\!\!\!\!\!\!\!\!\!\!\!\!\!\!\!  \text{subject to :}  \ \text{(\ref{eq : 14_1})},  \text{(\ref{eq : 14_2})},\text{(\ref{eq : 14_3})},\text{(\ref{eq : 14_4})}. \label{eq : 18_1}
\end{align}
\end{subequations}

It can be observed that after substituting the optimal value of  $\mathbf{\Gamma}$, as in (\ref{eq_11_1}), in equation (\ref{eq : 18}), the optimization problem in (\ref{eq : 18_m}) reduces exactly to the one in (\ref{eq : 14_m}). If we look at optimization problem (\ref{eq : 18_m}), we still have some non-linearities and non-convexities due to a ratio term and a product of two variables in the objective function. First, we attempt to handle the ratio term in  the objective function using the quadratic transformation approach \cite{shen2018fractional}. The quadratic transformation is performed on the objective function (\ref{eq : 18}) such that it becomes: 
\begin{align}
\begin{split}
\label{eq : 19}
\underset{\bm{\eta^{u}},\textbf{D}}{\max~} \ & \sum\limits_{t=1}^{T}\left( w\text{log}_2 (1+\Gamma_t)  -\alpha ||{\textbf{D}_t}||_{1}\right)  -\sum\limits_{t=1}^{T} u_{t}^2\left({A^2p_u\eta_{t}^{u}(\sum\limits_{m\in \mathcal{M}}d_{mt} \gamma_{mt})^2+{I}_t}\right) \\ & - \sum\limits_{t=1}^{T}\left(\Gamma_t - 2u_{t}\sqrt{w'(1+{\Gamma}_t)A^2p_u\eta_{t}^{u}(\sum\limits_{m\in \mathcal{M}} d_{mt}\gamma_{mt})^2}\right),
\end{split}
\end{align}
where $\textbf{u} = [u_1,u_2,...u_T]^\intercal$, $u_{t}$ is also the quadratic convex approximation parameter and need to be updated on the basis of solution of previous iteration.

The quadratic transformation in equation (\ref{eq : 19})  is the lower bound and concave approximation of ratio term $\frac{w'(1+{\Gamma}_t)A^2p_u\eta_{t}^{u}(\sum\limits_{m\in \mathcal{M}} d_{mt}\gamma_{mt})^2}{A^2p_u\eta_{t}^{u}(\sum\limits_{m\in \mathcal{M}}d_{mt} \gamma_{mt})^2+{I}_t}$ with respect to the parameter $u_t$. The optimal value of \textbf{u} can be calculated from equation (\ref{eq : 21_1}).
\begin{align}
\label{eq : 21_1}
& u^{*}_{t} \overset{\Delta}{=} \frac{\sqrt{w'(1+{\Gamma}_t)A^2p_u\eta_{t}^{u}(\sum\limits_{m\in \mathcal{M}}d_{mt} \gamma_{mt})^2}}{A^2p_u\eta_{t}^{u}(\sum\limits_{m\in \mathcal{M}} d_{mt}\gamma_{mt})^2+{I}_t}.
\end{align}
Therefore, equation (\ref{eq : 19}) is the lower bound to equation (\ref{eq : 18}), thus making the approximation feasible. 

The optimization problem becomes:
\begin{subequations}
\label{eq : 22_m}
\begin{align}
\label{eq : 22_1}
 \underset{\bm{\eta^{u}},\textbf{D}}{\max~} & \ \ \text{(\ref{eq : 19})} , \\
\text{subject to :} & \ \ \text{(\ref{eq : 14_1})},\text{(\ref{eq : 14_2})},\text{(\ref{eq : 14_3})},\text{(\ref{eq : 14_4})}.  \label{eq : 22_2}
\end{align}
\end{subequations}

After substituting the optimal value of \textbf{u} in equation (\ref{eq : 19}) and then solving optimization problem (\ref{eq : 22_m}) is equivalent to solving Problem (\ref{eq : 18_m}). Notice that Problem (\ref{eq : 22_m}) is a convex optimization problem in different blocks. If we fix $\textbf{D}$, then the objective function of problem (\ref{eq : 22_1}) becomes concave  with respect to $\bm{\eta^{u}}$. Then, the problem reduces to:
\begin{subequations}
\label{eq : 23_m}
\begin{align}
\label{eq : 23_1}
f_1 =  \underset{\bm{\eta^{u}}}{\max~} & \ \ \text{(\ref{eq : 19})} , \\
 \text{subject to :} & \ \ \text{(\ref{eq : 14_2})},\text{(\ref{eq : 14_3})}.  \label{eq : 23_2}
\end{align}
\end{subequations}
Similarly, upon fixing $\bm{\eta^{u}}$, the objective function of (\ref{eq : 22_m}) becomes concave with respect to $\textbf{D}$. Then, the optimization problem becomes:
\begin{subequations}
\label{eq : 24_m}
\begin{align}
\label{eq : 24_1}
f_2  = \underset{\textbf{D}}{\max~}& \ \ \text{(\ref{eq : 19})}, \\
 \text{subject to :} & \ \ \text{(\ref{eq : 14_1})},\text{(\ref{eq : 14_3})},\text{(\ref{eq : 14_4})}.  \label{eq : 24_2}
\end{align}
\end{subequations}
Problem \ref{eq : 22_m} can be solved following the proposed iterative approach as follows. This problem is decomposed into two distinct blocks, denoted as Problem \ref{eq : 23_m} and Problem \ref{eq : 24_m}. Initially, the solution for Problem \ref{eq : 23_m} is obtained while keeping $\textbf{D}$ fixed. Subsequently, based on the solution derived for Problem \ref{eq : 23_m}, the solution for Problem \ref{eq : 24_m} is determined. This is repeated until the convergence criterion is met. Algorithm \ref{Centralised algorithm 1} provides the details of this approach.
\begin{algorithm}[H]
\caption{Proposed Algorithm}\label{Centralised algorithm 1}
\begin{algorithmic}[1]
\STATE \textbf{Feasible Initialization:}  ${\bm{\eta^{u}}}^{(0)}$, $\textbf{D}^{(0)}$ and $i=0$, $\forall t$.
\REPEAT{}
\STATE $i=i+1$. 
\FORALL{t}
\STATE Compute  $\Gamma_{t}^{(i)}$  using  (\ref{eq_11_1}) and $u_t^{(i)}$ using (\ref{eq : 21_1}).
\ENDFOR
\STATE Solve optimization problem in (\ref{eq : 23_m}).

\FORALL{t}
\STATE Again compute $\Gamma_{t}^{(i)}$  using  (\ref{eq_11_1}) and $u_t^{(i)}$  using (\ref{eq : 21_1}).
\ENDFOR
\STATE Solve optimization problem in (\ref{eq : 24_m}).
\STATE Update ${\bm{\eta^{u}}}^{(i+1)} ={\bm{\eta^{u}}}^{(i)}$ and  $\textbf{D}^{(i+1)} =\textbf{D}^{(i)}$.
\UNTIL{$\left|\frac{f_{2}^{(i+1)}-f_{2}^{(i)}}{f_{2}^{(i)}}\right| \leq \epsilon$}
\end{algorithmic}
\end{algorithm}
\textit{Convergence Analysis} : 
For the convergence proof of the proposed Algorithm \ref{Centralised algorithm 1}, please refer to the appendix\ref{appendix}.

\textit{Complexity Analysis} : Problem \ref{eq : 23_m} has $T$ scalar variables, $3T$ linear constraints, so the complexity for solving Problem  \ref{eq : 23_m} is $\mathcal{O}(\sqrt{3T}4T^3)$ \cite{tam2016joint}. Similarly, the  complexity to solve Problem  \ref{eq : 24_m} with $MT$ scalar variables and $2(MT+T)$ linear constraint is give by $\mathcal{O}(\sqrt{2(MT+T)}(3MT+2T)(MT)^2)$ \cite{tam2016joint}.

\section{Performance Evaluation}
\label{simulations}
We perform numerical simulations on a $1000\times 1000$ square meters geographical area with APs and user equipment UEs randomly distributed. To simulate an infinite area network, a wrapped-around approach is used \cite{bjornson2020scalable}. Large-scale fading coefficients are modeled using a three-slope approach for path loss, incorporating an 8 dB standard deviation for shadow fading.  The minimum QoS parameter is taken as $0.2$ bits/s/Hz for each UE. Throughout our simulations, we keep  $T =40$, $M=100 \ \text{or} \ 150$, $A=4$, $L_c = 200$ and $L_p = 5$, while other parameters stay exactly the same as those outlined in \cite{ngo2017cell} or as explicitly stated. The penalty parameter, $\alpha$ is chosen by an ad-hoc approach. A more formal technique for selecting $\alpha$ will be discussed in our future work.
All numerical results are averaged over $100$ simulations. We do not compare our proposed approach with works in \cite{ngo2018performance} and \cite{guenach2020joint} as our proposed work focuses on the sum-throughput maximization whereas aforementioned works concentrate on maximizing the minimum Signal-to-Interference-Noise ratio. Furthermore, our proposed approach guarantees a minimum spectral efficiency of 0.2 bits/s/Hz for each UE, which was not explicitly addressed in above studies.

\begin{figure}[!h]
\centering
\includegraphics[width=0.9\textwidth]{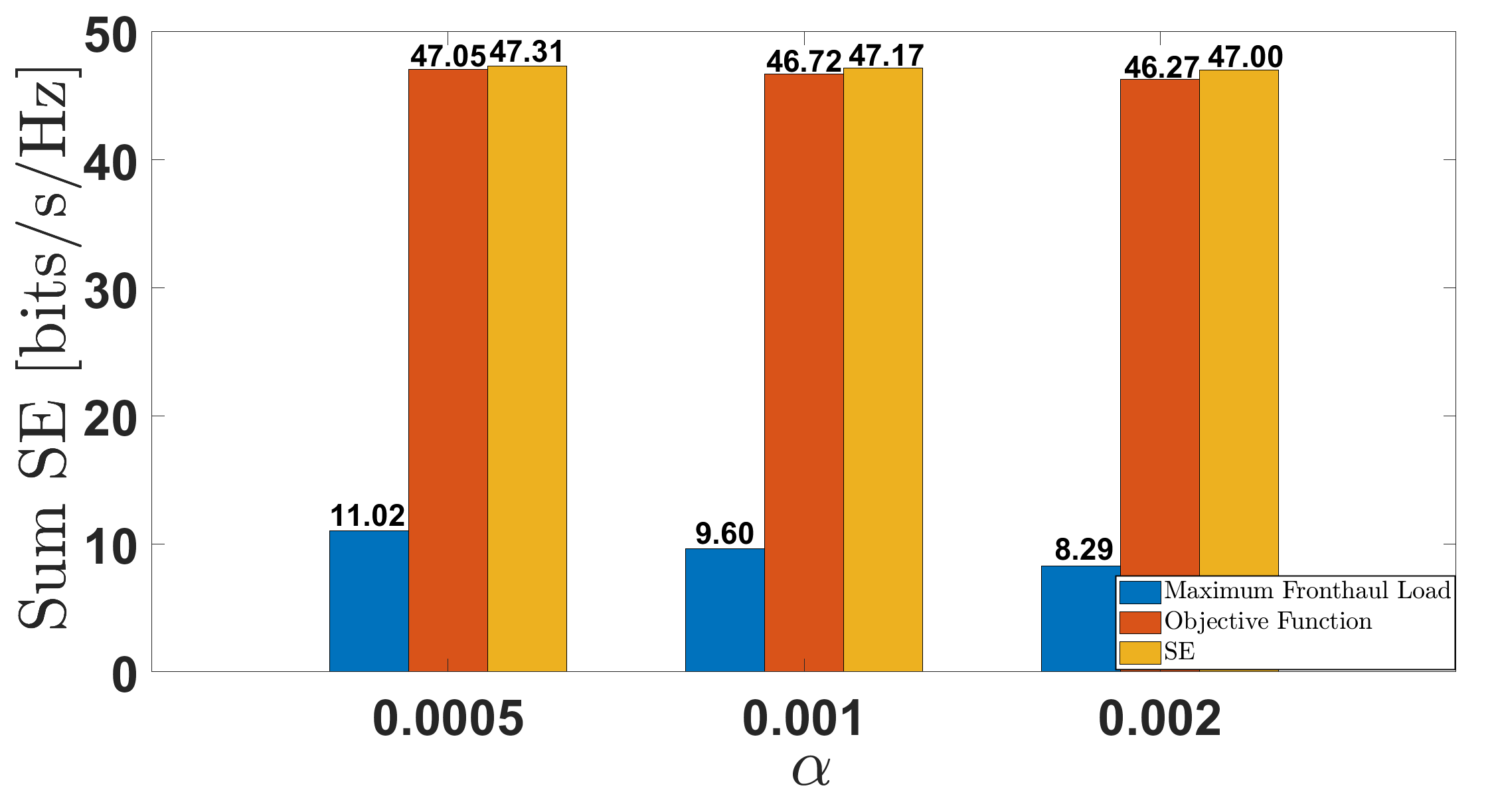}
\caption{The variation of maximum fronthaul load on an AP, objective function, and the sum SE, when $M=100$.}
\label{fig_2}
\end{figure}
\begin{figure}[!h]
\centering
\includegraphics[width=0.9\textwidth]{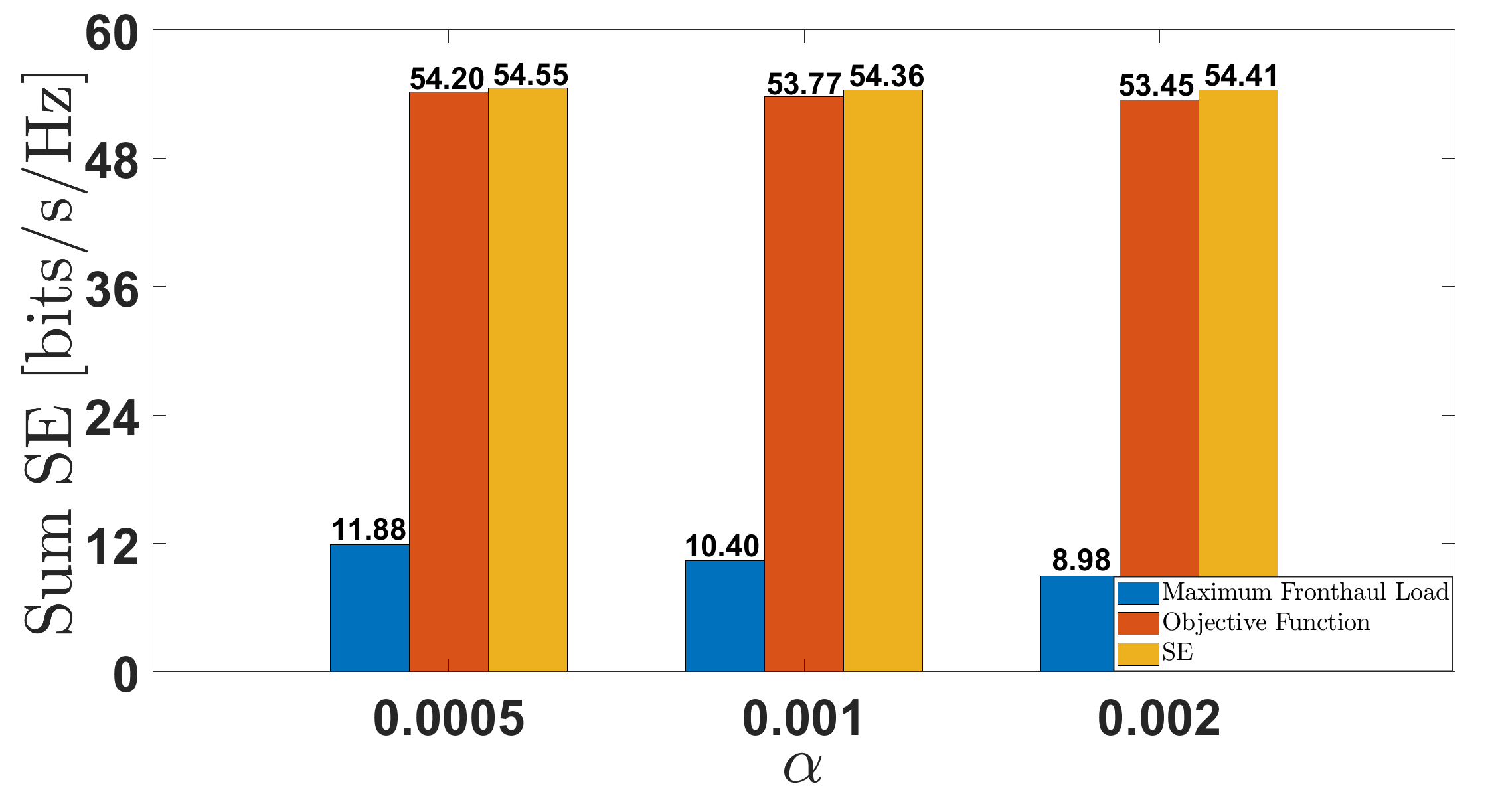}
\caption{The variation of maximum front-haul load on an AP, objective function, and the sum SE, when $M=150$.}
\label{fig_3}
\end{figure}

 
Fig. \ref{fig_2} shows the variations in maximum front-haul load, objective function, and the sum SE for three different regularization coefficient ($\alpha$) values, with the number of antennas fixed at $100$. This highlights how the penalty function affects the network's maximum front-haul load and total sum-throughput. Notably, the objective function value consistently remains lower than the total SE value, exhibiting a drop attributed to the penalty function. When $\alpha$ is increased by two and four times, the maximum front-haul load on an AP decreases by $13\%$ and $25\%$ respectively. This reduction is due to the increased $\alpha$, which limits AP-UE associations. As a consequence of the lesser number of AP-UE associations, there is a reduction in the sum SE by nearly $0.3\%$ and $0.6\%$ respectively. Similarly, Fig. \ref{fig_3} illustrates variability in maximum front-haul load, objective function, and sum SE for three different $\alpha$ values, however with $M$ set to $150$. For this scenario too, when the $\alpha$ is increased two and four times, the maximum front-haul load and total SE show a similar decreasing trend as in Fig. \ref{fig_2}. Therefore, the proposed approach effectively decreases the maximum front-haul load while not significantly reducing overall sum throughput, and this is achieved by a careful tuning of $\alpha$.
Examining Figs. \ref{fig_2} and \ref{fig_3}, we observe a trend that as the number of APs ($M$) increases from 100 to 150. The maximum front-haul load increases by $7.8\%$, $8.3\%$ and $8.4\%$ with regularization coefficients ($\alpha$) of 0.0005, 0.001 and 0.002, respectively. This growth can be attributed to the growing number of APs in the fixed area. As the AP count increases, there are more APs in the vicinity of each UE, resulting in a higher percentage of APs offering a strong channel or a larger large-scale fading coefficient for each UE. This gainful influence of more APs on the large-scale fading coefficient causes a change in the influence of the penalty function, allowing more APs to associate with each UE. While this increased association raises the maximum front-haul load, it also results in a significant ($15\%$) boost in the overall sum-throughput. In essence, the system responds to the increased AP count by increasing the strength of channels across numerous APs, hence increasing the network's overall throughput. The penalty function is a dynamic component of this process, optimizing AP-UE associations to optimize the objective function while reducing negative effects via penalties.

\begin{figure}[!h]
\centering
\includegraphics[width=0.9\textwidth]{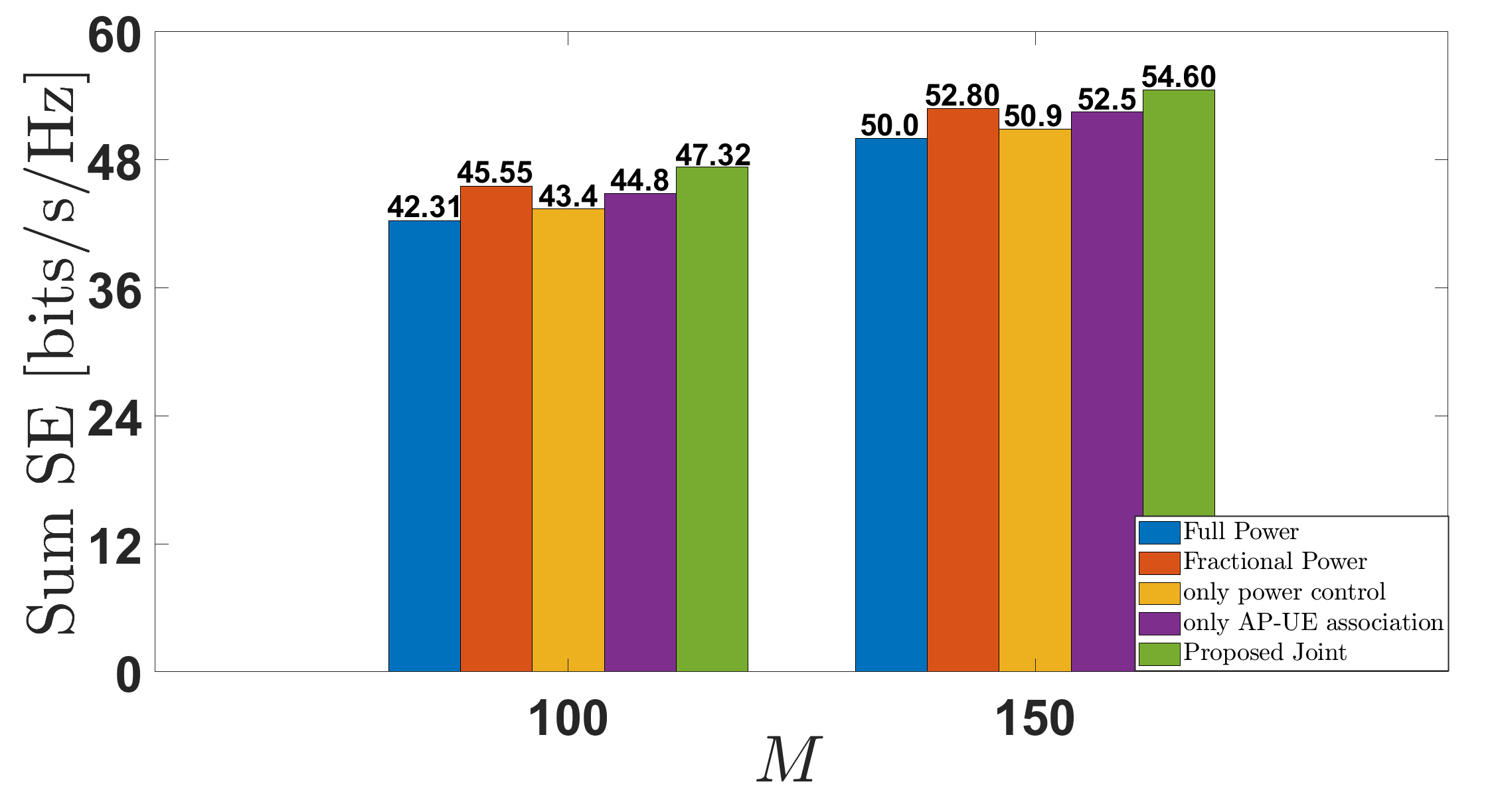}
\caption{The variation of the sum SE for different scenarios, when $\alpha =0.001$.}
\label{fig_4}
\end{figure}

\begin{figure}[!h]
\centering
\includegraphics[width=0.9\textwidth]{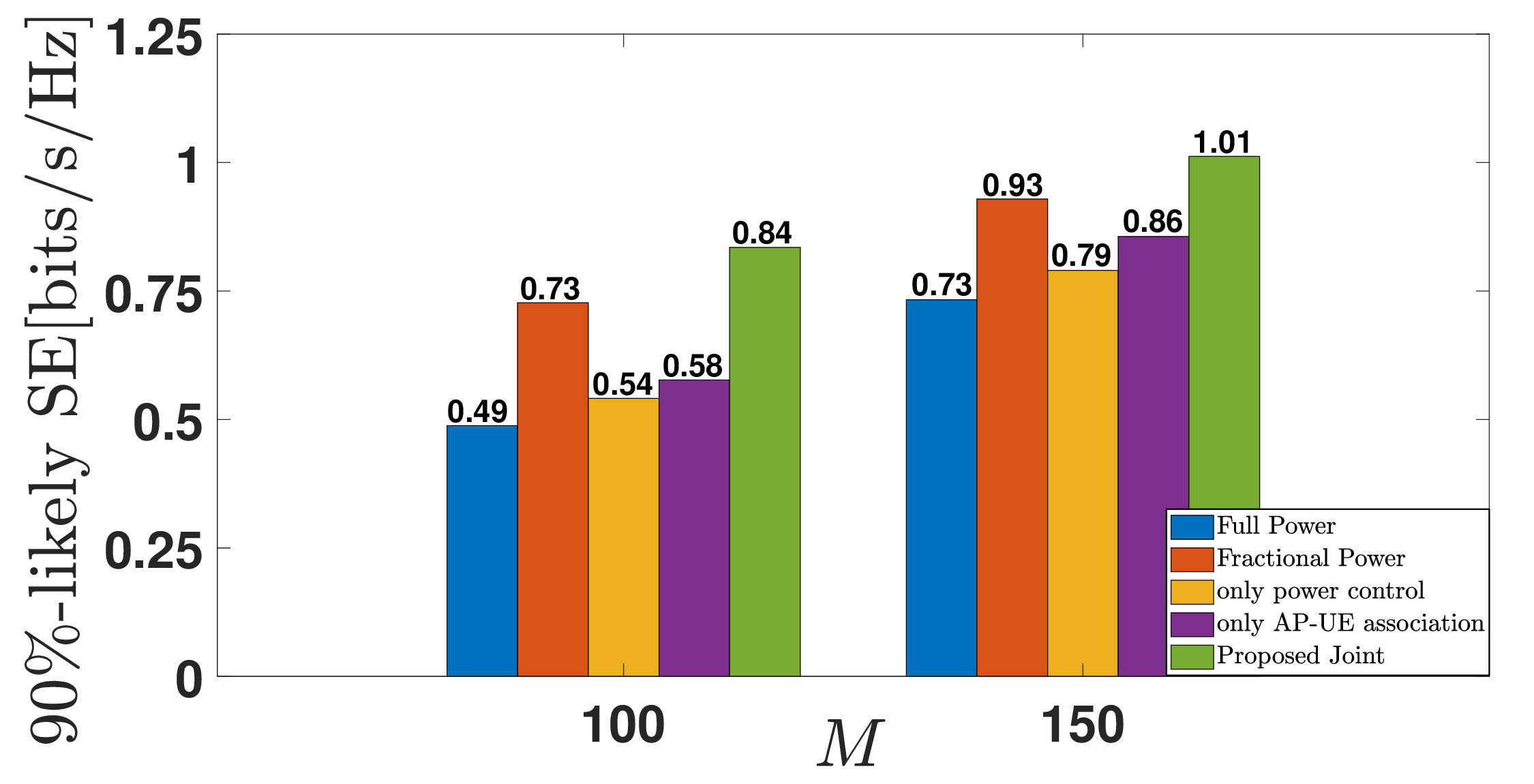}
\caption{CDF vs per-UE SE for different scenarios, when $\alpha =0.001$.}
\label{fig_5}
\end{figure}

Figs. \ref{fig_4} and  \ref{fig_5} illustrate the variation of sum SE and 90\%-likely SE respectively, of the proposed solution in comparison to the following four scenarios: a) where UEs transmit at full power and are served by all APs; b) fractional power control, when each AP serves all the UEs \cite{nikbakht2020uplink};  c) where only power control is performed using the proposed algorithm, when each AP serves all the UEs; and d) where only AP-UE association is performed using the proposed algorithm. 

Fig. \ref{fig_4} shows that when AP-UE association and power control are carried out simultaneously, the achievable sum SE is the largest. The proposed solution's sum SE surpasses that of scenarios `a', `b',`c', and `d' by $12\%$, $4\%$, $9\%$ and $5.5\%$, respectively, when $M=100$. Additionally, for case with $M=150$, the proposed solution outperforms these scenarios by $9\%$, $3.5\%$, $7\%$, and $4\%$, respectively. Our proposed approach outperforms the scenarios `a' and `b' as those do not meet the minimum QoS requirement. Moreover, in the scenarios `a', `b', and `c', the front-haul load can become impractical as all UEs may be served by all APs. This demonstrates the effectiveness of the proposed technique in optimizing the sum SE by jointly addressing the AP-UE association and power control problems.
Scenario 'b' performs better than scenario 'c' because scenario 'b' does not need to satisfy any constraints. Also scenario 'c' does not surpass proposed joint solution, despite UEs being served by all APs in scenario 'c'. As distant APs may have poor channel, those may negatively impact the SE of UEs, evidenced by the SE closed form (\ref{eq_11}). 

Indeed,  our proposed technique aims to enhance the sum SE. Fig. \ref{fig_5} shows how our strategy has a significant influence on the $90\%-likely$ per-user SE. Our proposed scheme outperforms scenarios `a', `b', `c', and `d' in terms of $90\%-likely$ per-user SE by $71\%$, $15\%$, $54\%$, and $45\%$, respectively, at $M=100$. Similarly, for $M=150$, our approach outperforms the same cases by $38\%$, $9\%$, $28\%$, and $18\%$ in terms of $90\%-likely$ per-user SE. The findings illustrate an additional benefit of the method we propose, demonstrating its massive contribution to improve per-user throughput while also increasing the total throughput.
\begin{figure}[!h]
\centering
\includegraphics[width=0.9\textwidth]{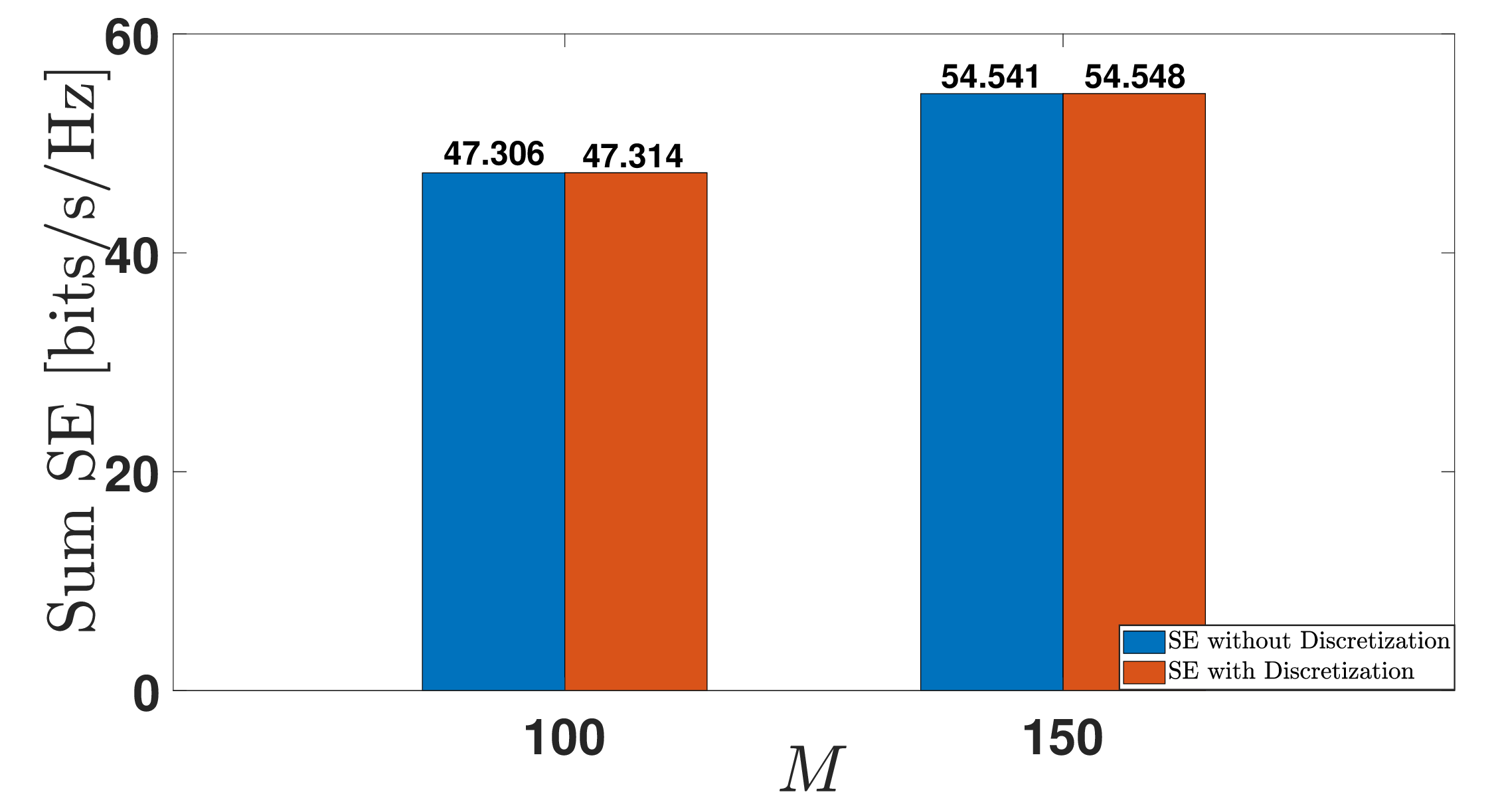}
\caption{The variation of SE with and without discretization of elements of  \textbf{D}.}
\label{fig_6}
\end{figure}  
In Fig. \ref{fig_6}, we provide a numerical justification for relaxation proposed in \eqref{eq : 14_9}. It shows the variation of the SE with and without the discretization of matrix \textbf{D}. The figure shows that for $M=100$ as well for $M=150$, the variations in the SE is negligible with and without the discretization of elements of matrix \textbf{D}. The negligible SE variations can be attributed to the fact that penalty parameter does not force elements of matrix \textbf{D} too much away from zero or one, thus our proposed solution does not require additional algorithm for AP-UE association. 
\begin{figure}[!h]
\centering
\includegraphics[width=0.9\textwidth]{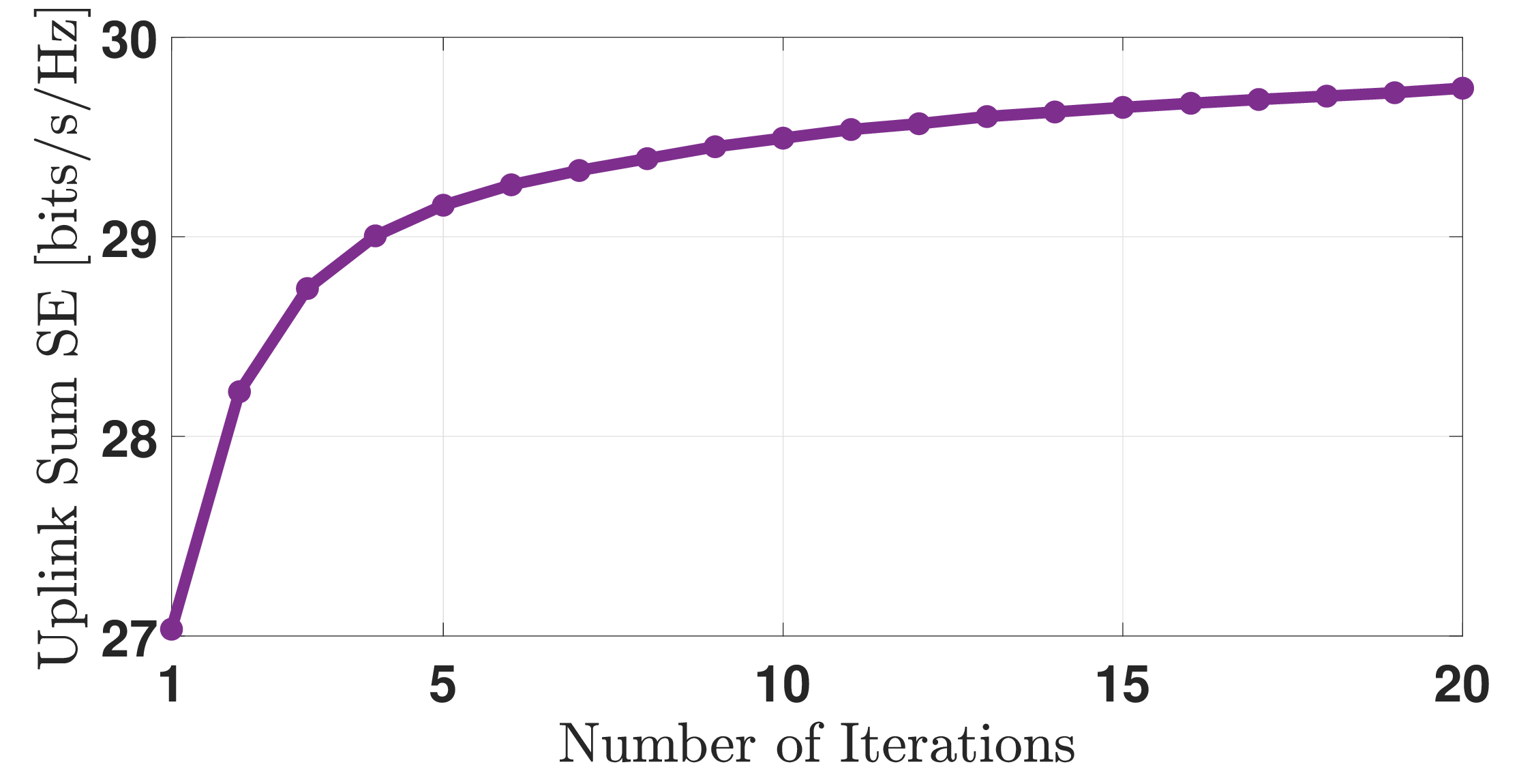}
\caption{The convergence graph of Algorithm \ref{Centralised algorithm 1}, when $\alpha = 0.001$, $T=30$ and $M=50$.}
\label{fig_1}
\end{figure} 

In Figure \ref{fig_1}, we provide a numerical evidence for the convergence Algorithm \ref{Centralised algorithm 1}. The figure shows that as the number of iterations increases, so does the value of the objective function, leading up in a non-decreasing function. However, the rate of rise gradually decreases, ensuring that the rate of increase of objective function finally becomes lower than the defined tolerance of $\epsilon=5 \times 10^{-3}$. This observation guarantees the convergence of Algorithm \ref{Centralised algorithm 1}.

\section{Conclusion}
\label{conclusion}
In this work, we address the challenge of joint AP-UE association and power control in uplink distributed mMIMO systems by introducing an \textit{l}1-penalty based joint mixed-integer non-convex optimization problem. Our technique strives to maximize system throughput while adhering to minimum quality of service (QoS) requirements. For tackling this optimization problem, we reformulate the original problem and develop an iterative technique that guarantees convergence. Through numerical simulations, we demonstrate that optimizing AP-UE association and power control factor jointly results in a significant increase in the system's sum-throughput when compared to the cases where just one parameter is optimized. Another interesting observation is the significant decrease in the maximum front-haul load at the expense of marginal drop in the sum-throughput. A highlight of our proposed scheme is the significant enhancement in the $90\%-likely$ per-UE SE. It demonstrates the significance and efficacy of the proposed scheme when compared to other schemes in maximizing system performance and providing a better user experience in uplink distributed mMIMO systems. Overall, our findings underline the need of optimizing AP-UE association and power regulation simultaneously, emphasizing the potential for significant throughput gains and per-user SE improvements in uplink distributed mMIMO systems.

\appendix
\label{appendix}
\textit{Convergence Analysis} :
For the convergence analysis of the Algorithm \ref{Centralised algorithm 1}, the primary goal is to prove that both optimization problem defined  in (\ref{eq : 23_m}) and (\ref{eq : 24_m}) converges to the their respective optimal solutions. Thereafter, we prove that in each iteration of Algorithm \ref{Centralised algorithm 1}, the  optimization function in (\ref{eq : 22_1}) is non-decreasing with respect to $\mathbf{\eta^u}$ and $\textbf{D}$. 

The feasible set of smooth concave objective function $\mathbb{F}_1$ (\ref{eq : 23_1}), articulated by Equation (\ref{eq : 23_2}), is closed and convex. As a result, $\mathbb{F}_1$ is inherently continuous and differentiable over its feasible set, implying the existence of an optimal solution for $\mathbb{F}_1$. The feasible set, defined by Equation (\ref{eq : 24_2}), of non-smooth concave objective function $\mathbb{F}_2$ (\ref{eq : 24_1}), is closed and convex.
However, $\mathbb{F}_2$ due to the \textit{l}1-penalty lacks continuous differentiability over its feasible set. Despite this, $\mathbb{F}_2$ can be separated into smooth and non-smooth part. The convergence of $\mathbb{F}_2$ is guaranteed if gradient of the smooth part of $\mathbb{F}_2$ is Lipschitz continuous over its feasible set \cite{scutari2016parallel}. Thus the second derivative of the $\mathbb{F}_2$ without the non-smooth part with respect to the AP-UE association variable, $d_{mt}$ is given by :
\begin{align}
\label{eq : 25}
 \frac{\partial^2\mathbb{F}_{1}}{\partial d_{mt}^{2}}= \sum\limits_{t=1}^{T}\left(2w'(1+\mathbf{\Gamma})A^2p_u\eta_{t}^{u}\gamma^{2}_{mt}    - \frac{P(\sum\limits_{t'=1}^{T}A^2p_u\eta_{t'}^{u}\left|\Psi^{H}_{t}\Psi_{t'} \right|^2(\gamma_{mt}\frac{\beta_{mt'}}{\beta_{mt}})^2)}{Q} - \frac{PQ'}{Q} +\frac{PQ'^2}{Q^2} \right),
\end{align}
where $P$, $Q$ and $Q'$ are given by, respectively:
\begin{align}
& P = w'(1+\mathbf{\Gamma})A^2p_u\eta_{t}^{u}(\sum\limits_{m\in \mathcal{M}} d_{mt}\gamma_{mt})^2,\\
& Q = \sum\limits_{t'=1}^{T}A^2p_u\eta_{t'}^{u}\left|\Psi^{H}_{t}\Psi_{t'} \right|^2(\sum\limits_{m\in \mathcal{M}}d_{mt}\gamma_{mt}\frac{\beta_{mt'}}{\beta_{mt}})^2+\sum\limits_{t'=1}^{T}Ap_u\eta_{t'}^{u}\sum\limits_{m\in \mathcal{M}}d_{mt}\gamma_{mt}\beta_{mt'} + \sum\limits_{m\in \mathcal{M}}Ad_{mt}\gamma_{mt},\\
& Q' = \sum\limits_{t'=1}^{T}A^2p_u\eta_{t'}^{u}\left|\Psi^{H}_{t}\Psi_{t'} \right|^2(\sum\limits_{m\in \mathcal{M}}d_{mt}\gamma_{mt}\frac{\beta_{mt'}}{\beta_{mt}})\gamma_{mt}\frac{\beta_{mt'}}{\beta_{mt}}+\sum\limits_{t'=1}^{T}Ap_u\eta_{t'}^{u}\gamma_{mt}\beta_{mt'} +A\gamma_{mt}
\end{align}

Notably, all the terms in the above double  partial derivative (\ref{eq : 25}) are bounded from above, thus  the gradient of the $\mathbb{F}_2$ is Lipschitz continuous over its feasible set.   

To  prove the non-decreasing nature of the optimization function $f(\mathbf{\eta^u},\textbf{D})$ defined in (\ref{eq : 22_1}), let ${\mathbf{\eta^u}}^*$ be the optimal value of function $f$, when $\textbf{D}$ is fixed. Then, for fixed $\textbf{D}$, the inequality $f({\mathbf{\eta^u}}^*,{\textbf{D}}^{(i)}) \geq f({\mathbf{\eta^u}}^{(i)},{\textbf{D}}^{(i)})$ always holds due to the  concavity of the function $f$. For the optimal  $\textbf{D}^*$, when $\mathbf{\eta^u}$ is fixed at ${\mathbf{\eta^u}}^{(i+1)}$,  the inequality $f({\mathbf{\eta^u}}^{(i+1)},{\textbf{D}}^{*}) \geq f({\mathbf{\eta^u}}^{(i+1)},{\textbf{D}}^{(i)})$ always holds as $f$ is concave for fixed $\textbf{D}$. It can be inferred  that  $f({\mathbf{\eta^u}}^{(i+1)},{\textbf{D}}^{(i+1)}) \geq f({\mathbf{\eta^u}}^{(i)},{\textbf{D}}^{(i)})$, indicating $f$ is  non-decreasing in each iteration. It makes the optimization function monotonically increasing in each iteration and also the optimization function is bounded from above, thereby guaranteeing the convergence of the proposed algorithm.

\end{document}